\documentclass[lettersize,journal]{IEEEtran}
\usepackage{amsmath,amsfonts}
\usepackage{algorithmic}
\usepackage{algorithm}
\usepackage{array}
\usepackage[caption=false,font=footnotesize,labelfont=sf,textfont=sf]{subfig}
\usepackage{textcomp}
\usepackage{stfloats}
\usepackage{url}
\usepackage{verbatim}
\usepackage{graphicx}
\usepackage{cite}
\usepackage{amssymb}
\usepackage{titlesec}
\usepackage{enumitem}
\usepackage{color}

\begin{document}
\title{RIS-Assisted Simultaneous Legitimate Monitoring and Jamming for Industrial Wireless Networks}

\author{Likang Zhang, \IEEEmembership{Graduate Student Member, IEEE}, Qinghe Du, \IEEEmembership{Member, IEEE}, \\ Yijing Ren, \IEEEmembership{Graduate Student Member, IEEE}, Xiao Tang, \IEEEmembership{Member, IEEE}, \\Maged Elkashlan, \IEEEmembership{Senior Member, IEEE}, and Zhu Han, \IEEEmembership{Fellow, IEEE}

\thanks{This work was supported in part by Key Research and Development Program of Shaanxi Province under Grant No. 2023-GHZD-05, National Natural Science Foundation of China under Grant No.~62071373, and Innovation Capability Support Program of Shaanxi under Grant No. 2021TD-08.}
\thanks{Likang Zhang, Qinghe Du (Correspondence author) and Xiao Tang are with the School of Information and Communications Engineering, Xi’an Jiaotong University, China (e-mails: zhanglk20000306@stu.xjtu.edu.cn; duqinghe@mail.xjtu.edu.cn; tangxiao@xjtu.edu.cn).}
\thanks{Yijing Ren is with the Department of Engineering, King's College London, U.K. (e-mail: yijing.ren@kcl.ac.uk).}
\thanks{Maged Elkashlan is with the School of Electronic Engineering and Computer
Science, Queen Mary University of London, E1 4NS London, U.K.
(e-mail: maged.elkashlan@qmul.ac.uk).}
\thanks{Zhu Han is with the Department of Electrical and Computer Engineering, University of Houston, Houston, TX 77004 USA, and also with the Department of Computer Science and Engineering, Kyung Hee University, Seoul 446-701, South Korea (e-mail: hanzhu22@gmail.com).}
}


\IEEEpubid{0000--0000/00\$00.00~\copyright~2021 IEEE}

\maketitle
\begin{abstract}
In this paper, we study reconfigurable intelligent surface (RIS)-assisted simultaneous legitimate monitoring and jamming techniques for industrial environments, so that legitimate monitor (LM) and legitimate jammers (LJs) can sustainably monitor and interfere with suspicious communications with minimum transmission power. 
Specifically, we propose a Block Coordinate Descent-Particle Swarm Optimization (BCD-PSO) based scheme to optimize RIS's phase shift matrix and minimize LJs’ transmission power, while successfully jamming and stably monitoring unauthorized communications. 
Simulation results demonstrate that the proposed BCD-PSO can enhance the performances in terms of monitoring, resource utilization and robustness. Moreover, we effectively exam the best deployment of RIS towards diverse objectives.
\end{abstract}


\section{Introduction}

\IEEEPARstart{R}{econfigurable} intelligent surface (RIS) is recognized as one of the key technologies towards 6G. By adjusting phases of reflective elements, propagation of electromagnetic waves can be tuned to reshape communications environments. It can not only increase sum rate and optimize resource utilization \cite{10128963,10061580}, but also significantly enhance information security \cite{9446526}. The appealing advantages of RIS have led to extensive researches focused on channel modeling, channel estimation, transmission design, localization techniques, and prototype design, etc, thus resulting in better performance \cite{pan2022overview}.

In recent years, various researches have been developed to tackle malicious attacks, such as channel hopping, power control and adaptive nulling antennas. However, the disadvantages including waste of spectrum resources, constraint of power and high complexity cannot be ignored \cite{9672153}. To overcome shortcomings and enhance information security, RIS have been introduced.
The authors of \cite{10066528} divided RIS for serving legitimate users and interfering illegitimate users, where partition ratio and transmission power were optimized.
In \cite{9685957}, a RIS-assisted scenario against eavesdroppers was considered, and a block coordinate ascent method was adopted to maximize secrecy rate.
The authors of \cite{9262884} applied artificial noise (AN) in RIS-assisted system to degrade eavesdropper's performance while improving secrecy rate.
In \cite{9672153}, a RIS-assisted scenario was investigated to minimize transmit power subject to outage probability and signal to interference plus noise ratio (SINR). It also examined that the best deployment of RIS should be close to the base station.

While existing researches mainly concentrated on security protection for legitimate users, techniques against unauthorized users recently has been explored \cite{9417411,9797949,hu2022analysis}. 
The authors of \cite{9797949} proposed an monitoring scheme for illegitimate users with Majorization-Minimization and Alternating Direction Method of Multipliers algorithms. However, it cannot interrupt suspicious communication.
In \cite{hu2022analysis}, a RIS-assisted eavesdropping and jamming strategy was presented. However, the monitoring and interference cannot be guaranteed simultaneously in this work. 
To address these shortcomings, we innovatively study the simultaneous legitimate monitoring and jamming techniques to block while intercepting the illegitimate information.

\IEEEpubidadjcol 

Automated manufacturing processes increasingly rely on wireless technologies that connect sensors, robots and controlling systems, which also bring convenience for illegal users due to the broadcast nature of wireless channel. 
For example, a suspicious transmitter (ST) delivers manufacturing information to a suspicious receiver (SR). Both ST and SR are hacked, thus system cannot directly turn them off. Therefore, there are urgent needs to develop legitimate monitoring with jamming techniques, to block suspicious communications while identifying what information are being targeted, to strengthen the protection of specific information. Following these objectives, the contributions of this paper are summarized as follows:

\begin{itemize}
\item{Application Scenario: We present an early research with simultaneous monitoring and jamming suspicious communications in RIS-assisted system for industrial wireless networks, which consists of two illegitimate devices (ST and SR), a collaborative monitoring and jamming system (LM and LJs), and a RIS which assists LM and LJs.}
\item{Scheme Design: We propose a Block Coordinate Descent-Particle Swarm Optimization (BCD-PSO) based scheme, which can minimize the transmission power of LJs, jam unauthorized communications, and intercept suspicious information by optimizing RIS's phase shift matrix.}
\item{Performance Enhancement: By using BCD-PSO, LJs and LM can achieve better performances in monitoring, transmission power and robustness. We also effectively exam the best deployment of RIS towards different objectives.}
\end{itemize}

The remainder of the paper is organized as follows. Sections \ref{S2} and \ref{S3} introduce model and problem formulation of RIS-assisted legitimate monitoring and jamming system. Then, we propose BCD-PSO in Section \ref{S4}. Section \ref{S5} presents simulation results. Finally, the paper concludes in Section \ref{S6}.

\section{System Model} \label{S2}
\begin{figure}[!t]
\vspace*{-0.2\baselineskip}
\centering
\includegraphics[width=3.4in]{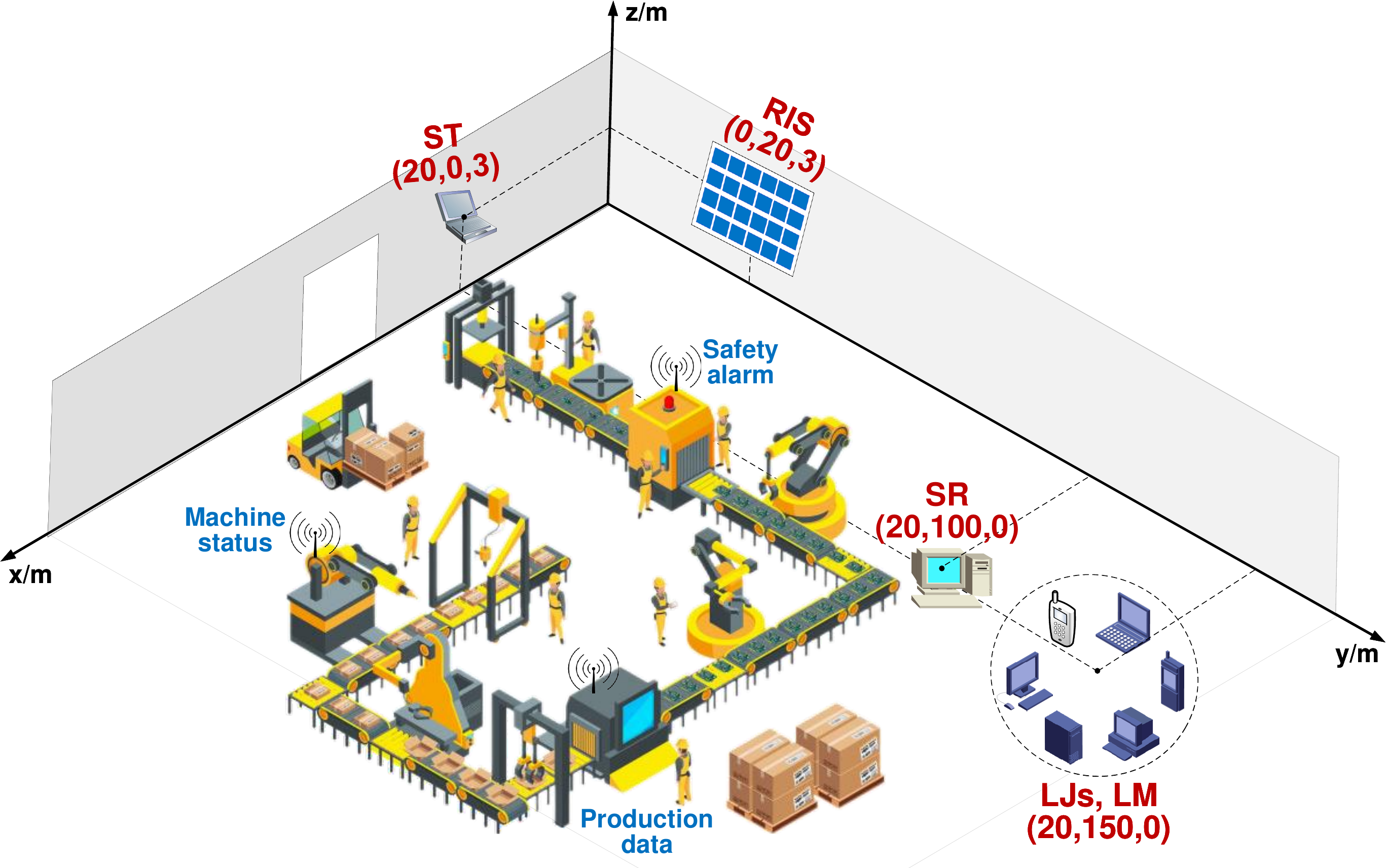}
\caption{The model of RIS-assisted legitimate monitoring and jamming system.}
\label{figure1}
\vspace*{-0.4\baselineskip}
\end{figure}

We consider a RIS-assisted legitimate monitoring and jamming system for industrial environments. As shown in Fig.~\ref{figure1}, there is a single-antenna SR and a ST with $M_\mathrm{t}$ antennas. We deploy a RIS with $N$ elements, assisting a single-antenna LM and $K$ single-antenna LJs.
We assume that suspicious communications between ST and SR have been detected. However, both ST and SR are hacked, thus the system cannot turn them off directly. Therefore, LJs send AN to jam SR and meanwhile LM keeps monitoring to identify the targeted information.
The channels of ST-to-RIS, RIS-to-SR, LJs-to-RIS, RIS-to-LM, ST-to-LM, ST-to-SR, and LJs-to-SR links are denoted as $\mathbf{H}_\mathrm{TI} \in \mathbb{C}^{M_\mathrm{t} \times N}$, $\mathbf{h}_\mathrm{IR} \in \mathbb{C}^{N \times 1}$, $\mathbf{H}_\mathrm{KI} \in \mathbb{C}^{N \times K}$, $\mathbf{h}_\mathrm{IM} \in \mathbb{C}^{N \times 1}$, $\mathbf{h}_\mathrm{TM} \in \mathbb{C}^{1 \times M_\mathrm{t}}$, $\mathbf{h}_\mathrm{TR} \in \mathbb{C}^{1 \times M_\mathrm{t}}$, and $\mathbf{h}_\mathrm{KR} \in \mathbb{C}^{1 \times K}$, respectively, where $\mathbb{C}$ denotes the complex number set.
In recent works, there are several channel estimation techniques in RIS-assisted systems \cite{9779086,10333823}, based on which we assume that all channel state information (CSI) is known to LJs and LM. The RIS's phase shift matrix is denoted by $\mathbf{\Phi} \triangleq \mathrm{diag}( [e^{j\theta_1} ~ e^{j\theta_2} ~ \ldots ~ e^{j\theta_N}])$, where $\theta_n$ is the phase shift of the $n$th element. In addition, due to severe path losses, we ignore signals reflected twice or more by RIS.

\section{Problem Formulation} \label{S3}
In this system, ST uses maximum ratio transmission (MRT) strategy to maximize strength of signal. The LJs also adopt MRT, which achieves lower complexity with a concentrated signal to interfere with SR as soon as possible and reduce leakage of information.
The beamforming matrices of ST and LJs, denoted by $\mathbf{w}_\mathrm{ST}$ and $\mathbf{w}_\mathrm{J}$, respectively, are given by
\begin{eqnarray}
    \left\{
        \begin{array}{lcl}
             \mathbf{w}_\mathrm{ST} = \frac{\sqrt{P_\mathrm{ST}} \mathbf{h}_\mathrm{TR}^\mathrm{H} }{\|\mathbf{h}_\mathrm{TR}^\mathrm{H} \|}, \\
             \mathbf{w}_\mathrm{J} = \frac{ \sqrt{P_\mathrm{J}}( \mathbf{h}_\mathrm{KR}+\mathbf{h}_\mathrm{TR}^\mathrm{H}\mathbf{\Phi}\mathbf{H}_\mathrm{KI}) }{\|\mathbf{h}_\mathrm{KR}+\mathbf{h}_\mathrm{TR}^\mathrm{H}\mathbf{\Phi}\mathbf{H}_\mathrm{KI} \|}.
        \end{array}
    \right.
\end{eqnarray}
Here, $P_\mathrm{ST}$ and $P_\mathrm{J}$ denote transmission power of ST and LJs, respectively. Considering cooperation between LM and LJs, the impact of AN on LM can be ignored because LM can remove it. Therefore, signals received by SR and LM, denoted by $\mathbf{y}_\mathrm{SR}$ and $\mathbf{y}_\mathrm{M}$, respectively, can be written as
\begin{eqnarray}
    \left\{
        \begin{array}{lcl}
            \mathbf{y}_\mathrm{SR} = \mathbf{h}_\mathrm{s} \mathbf{w}_\mathrm{ST} x_\mathrm{ST} + \mathbf{h}_\mathrm{i} \mathbf{w}_\mathrm{J}^\mathrm{H} x_\mathrm{J} + \mathbf{n}_\mathrm{SR}, \\
            \mathbf{y}_\mathrm{M} = \mathbf{h}_\mathrm{m} \mathbf{w}_\mathrm{ST} x_\mathrm{ST} + \mathbf{n}_\mathrm{M},
        \end{array}
    \right.
\end{eqnarray}
where $\mathbf{h}_\mathrm{s}\triangleq \mathbf{h}_\mathrm{TR}+\mathbf{h}_\mathrm{IR}^\mathrm{H}\mathbf{\Phi}\mathbf{H}_\mathrm{TI}^\mathrm{H}$, $\mathbf{h}_\mathrm{i}\triangleq \mathbf{h}_\mathrm{KR}+\mathbf{h}_\mathrm{IR}^\mathrm{H}\mathbf{\Phi}\mathbf{H}_\mathrm{KI}$, and $\mathbf{h}_\mathrm{m}\triangleq\mathbf{h}_\mathrm{TM}+\mathbf{h}_\mathrm{IM}^\mathrm{H} \mathbf{\Phi} \mathbf{H}_\mathrm{TI}^\mathrm{H}$. Moreover, $x_\mathrm{ST}$ and $x_\mathrm{J}$ are signals transmitted by ST and LJs, respectively, where $\mathbb{E}\{|x_\mathrm{ST}|^2\}=\mathbb{E}\{|x_\mathrm{J}|^2\} = 1$. In addition, $\mathbf{n}_\mathrm{SR} \sim \mathcal{CN}(0,\sigma_\mathrm{SR}^2)$ and $\mathbf{n}_\mathrm{M} \sim \mathcal{CN}(0,\sigma_\mathrm{M}^2)$ both are additive white Gaussian noises. Then, the SINR $\gamma_\mathrm{SR}$ of SR and the SNR $\gamma_\mathrm{M}$ of LM are given by
\begin{eqnarray} \label{eq1}
    \left\{
        \begin{array}{lcl}
            \gamma_\mathrm{SR} = \frac{\|(\mathbf{h}_\mathrm{TR}+\mathbf{h}_\mathrm{IR}^\mathrm{H}\mathbf{\Phi}\mathbf{H}_\mathrm{TI}^\mathrm{H})\mathbf{w}_\mathrm{ST} x_\mathrm{ST}\|^2}{\|(\mathbf{h}_\mathrm{KR}+\mathbf{h}_\mathrm{IR}^\mathrm{H}\mathbf{\Phi}\mathbf{H}_\mathrm{KI})\mathbf{w}_\mathrm{J}^\mathrm{H} x_\mathrm{J}\|^2+\sigma_\mathrm{SR}^2},  \\
            \gamma_\mathrm{M} = \frac{1}{\sigma_\mathrm{M}^2}\|(\mathbf{h}_\mathrm{TM}+\mathbf{h}_\mathrm{IM}^\mathrm{H} \mathbf{\Phi} \mathbf{H}_\mathrm{TI}^\mathrm{H}) \mathbf{w}_\mathrm{ST} x_\mathrm{ST}\|^2.
        \end{array}
    \right.
\end{eqnarray}

To efficiently jam and intercept the suspicious information, we jointly optimize RIS's phase shift matrix and minimize LJs' transmission power to confine SR's SINR less than a specified threshold $\gamma_\mathrm{SR,th}$ while maintaining LM's SNR beyond $\gamma_\mathrm{M,th}$. Therefore, the optimization problem can be formulated as:
\begin{align}
    \boldsymbol{P1} \ \min_{\mathbf{\Phi}} \ & P_\mathrm{J} \\
    \text{s.t.:} \ &\frac{\|(\mathbf{h}_\mathrm{TR}+\mathbf{h}_\mathrm{IR}^\mathrm{H}\mathbf{\Phi}\mathbf{H}_\mathrm{TI}^\mathrm{H})\mathbf{w}_\mathrm{ST}\|^2}{\|(\mathbf{h}_\mathrm{KR}+\mathbf{h}_\mathrm{IR}^\mathrm{H}\mathbf{\Phi}\mathbf{H}_\mathrm{KI})\mathbf{w}_\mathrm{J}^\mathrm{H}\|^2+\sigma_\mathrm{SR}^2} 
    \leq \gamma_\mathrm{SR,th}; \label{eq3} \\ 
    & \frac{1}{\sigma_\mathrm{M}^2}\|(\mathbf{h}_\mathrm{TM}+\mathbf{h}_\mathrm{IM}^\mathrm{H} \mathbf{\Phi} \mathbf{H}_\mathrm{TI}^\mathrm{H}) \mathbf{w}_\mathrm{ST}\|^2 \geq \gamma_\mathrm{M,th}; \label{eq11} \\
    & |v_n|^2=1, \ n=1,\ldots, N  \label{eq12}, 
\end{align}
where $v_n=e^{j\theta_n}$. Considering non-convexity of $\boldsymbol{P1}$, the closed-form of optimal solution is intractable. Then, we propose an efficient scheme based on BCD and PSO algorithms.

\section{Block Coordinate Descent-Particle Swarm Optimization based Scheme} \label{S4}
\subsection{Problem Converting}
We first convert the problem to a manageable form. According to (\ref{eq3}), we have
\begin{equation} \label{eq9}
    P_\mathrm{J} \geq \frac{\|(\mathbf{h}_\mathrm{TR}+\mathbf{h}_\mathrm{IR}^\mathrm{H}\mathbf{\Phi}\mathbf{H}_\mathrm{TI}^\mathrm{H})\mathbf{w}_\mathrm{ST}\|^2-\sigma_\mathrm{SR}^2 \gamma_\mathrm{SR,th}}{\gamma_\mathrm{SR,th}\| (\mathbf{h}_\mathrm{KR}+\mathbf{h}_\mathrm{IR}^\mathrm{H}\mathbf{\Phi}\mathbf{H}_\mathrm{KI} )\overline{\mathbf{w}}_\mathrm{J}^\mathrm{H}\|^2},
\end{equation}
where $\overline{\mathbf{w}}_\mathrm{J} \triangleq \frac{\mathbf{h}_\mathrm{KR}+\mathbf{h}_\mathrm{IR}^\mathrm{H}\mathbf{\Phi}\mathbf{H}_\mathrm{KI}}{\|\mathbf{h}_\mathrm{KR}+\mathbf{h}_\mathrm{IR}^\mathrm{H}\mathbf{\Phi}\mathbf{H}_\mathrm{KI}\|}$. Clearly,  (\ref{eq9}) gives a approximate optimum of $P_\mathrm{J}$. Next, we aim to minimize the approximate optimum of $P_\mathrm{J}$, which can be written as
\begin{align}
    \boldsymbol{P2} \ \min_{\mathbf{\Phi}} & \ \frac{\|(\mathbf{h}_\mathrm{TR}+\mathbf{h}_\mathrm{IR}^\mathrm{H}\mathbf{\Phi}\mathbf{H}_\mathrm{TI}^\mathrm{H})\mathbf{w}_\mathrm{ST}\|^2-\sigma_\mathrm{SR}^2 \gamma_\mathrm{SR,th}}{\gamma_\mathrm{SR,th}\| (\mathbf{h}_\mathrm{KR}+\mathbf{h}_\mathrm{IR}^\mathrm{H}\mathbf{\Phi}\mathbf{H}_\mathrm{KI} )\overline{\mathbf{w}}_\mathrm{J}^\mathrm{H}\|^2}   \label{eq4} \\
    \text{s.t.:} & \ \text{ (\ref{eq11}) and  (\ref{eq12})}.
\end{align}
The numerator and denominator of  (\ref{eq4}) and the left-hand side of  (\ref{eq11}) can be rewritten as
\begin{eqnarray}
    \left\{
        \begin{array}{lcl}
            \|(\mathbf{h}_\mathrm{TR}+\mathbf{h}_\mathrm{IR}^\mathrm{H}\mathbf{\Phi}\mathbf{H}_\mathrm{TI}^\mathrm{H})\mathbf{w}_\mathrm{ST}\|^2-\sigma_\mathrm{SR}^2 \gamma_\mathrm{SR,th} \nonumber \\
            \quad\quad  = \mathbf{v}\mathbf{A}_\mathrm{ST}\mathbf{v}^\mathrm{H} + \mathbf{v}\mathbf{b}_\mathrm{ST} + \mathbf{b}_\mathrm{ST}^\mathrm{H}\mathbf{v}^\mathrm{H} + c_\mathrm{ST} -\sigma_\mathrm{SR}^2 \gamma_\mathrm{SR,th}, \\
            \gamma_\mathrm{SR,th}\| (\mathbf{h}_\mathrm{KR} +\mathbf{h}_\mathrm{IR}^\mathrm{H}\mathbf{\Phi}\mathbf{H}_\mathrm{KI} )\overline{\mathbf{w}}_\mathrm{J}^\mathrm{H}\|^2 \\
            \quad\quad  = \gamma_\mathrm{SR,th}(\mathbf{v}\mathbf{A}_\mathrm{J}\mathbf{v}^\mathrm{H} + \mathbf{v}\mathbf{b}_\mathrm{J} + \mathbf{b}_\mathrm{J}^\mathrm{H}\mathbf{v}^\mathrm{H} + c_\mathrm{J} ),  \\
            \frac{1}{\sigma_\mathrm{M}^2}\|(\mathbf{h}_\mathrm{TM}+\mathbf{h}_\mathrm{IM}^\mathrm{H}\mathbf{\Phi}\mathbf{H}_\mathrm{TI}^\mathrm{H})\mathbf{w}_\mathrm{ST}\|^2 \\
            \quad\quad  = \frac{1}{\sigma_\mathrm{M}^2}(\mathbf{v}\mathbf{A}_\mathrm{M}\mathbf{v}^\mathrm{H} + \mathbf{v}\mathbf{b}_\mathrm{M} + \mathbf{b}_\mathrm{M}^\mathrm{H}\mathbf{v}^\mathrm{H} + c_\mathrm{M}),
        \end{array}
    \right.
\end{eqnarray}
where $\mathbf{v} \! \triangleq \! [v_1 ~ v_2 ~ \ldots ~ v_N]$, 
$\mathbf{A}_\mathrm{ST} \! \triangleq \! \mathbf{a}_\mathrm{ST}\mathbf{a}_\mathrm{ST}^\mathrm{H}$, 
$\mathbf{a}_\mathrm{ST} \! \triangleq \! \mathrm{diag}(\mathbf{h}_\mathrm{IR}^\mathrm{H})\mathbf{H}_\mathrm{TI}^\mathrm{H}\mathbf{w}_\mathrm{ST}$, 
$\mathbf{b}_\mathrm{ST} \! \triangleq \! \mathrm{diag}(\mathbf{h}_\mathrm{IR}^\mathrm{H})\mathbf{H}_\mathrm{TI}^\mathrm{H}\mathbf{w}_\mathrm{ST}\mathbf{w}_\mathrm{ST}^\mathrm{H}\mathbf{h}_\mathrm{TR}^\mathrm{H}$, 
$c_\mathrm{ST} \! \triangleq \! \mathbf{h}_\mathrm{TR}\mathbf{w}_\mathrm{ST}\mathbf{w}_\mathrm{ST}^\mathrm{H}\mathbf{h}_\mathrm{TR}^\mathrm{H}$, 
$\mathbf{A}_\mathrm{J} \! \triangleq \! \mathbf{a}_\mathrm{J}\mathbf{a}_\mathrm{J}^\mathrm{H}$, 
$\mathbf{a}_\mathrm{J} \! \triangleq \! \mathrm{diag}(\mathbf{h}_\mathrm{IR}^\mathrm{H})\mathbf{H}_\mathrm{KI}\overline{\mathbf{w}}_\mathrm{J}^\mathrm{H}$,  
$\mathbf{b}_\mathrm{J} \! \triangleq \! \mathrm{diag}(\mathbf{h}_\mathrm{IR}^\mathrm{H})\mathbf{H}_\mathrm{KI}\overline{\mathbf{w}}_\mathrm{J}^\mathrm{H}\overline{\mathbf{w}}_\mathrm{J}\mathbf{h}_\mathrm{KR}^\mathrm{H}$, 
$c_\mathrm{J} \! \triangleq \! \mathbf{h}_\mathrm{KR}\overline{\mathbf{w}}_\mathrm{J}^\mathrm{H}\overline{\mathbf{w}}_\mathrm{J}\mathbf{h}_\mathrm{KR}^\mathrm{H}$, 
$\mathbf{A}_\mathrm{M} \! \triangleq \! \mathbf{a}_\mathrm{M}\mathbf{a}_\mathrm{M}^\mathrm{H}$, 
$\mathbf{a}_\mathrm{M} \! \triangleq \! \mathrm{diag}(\mathbf{h}_\mathrm{IM}^\mathrm{H})\mathbf{H}_\mathrm{TI}^\mathrm{H}\mathbf{w}_\mathrm{ST}$,
$\mathbf{b}_\mathrm{M} \! \triangleq \! \mathrm{diag}(\mathbf{h}_\mathrm{IM}^\mathrm{H})\mathbf{H}_\mathrm{TI}^\mathrm{H}\mathbf{w}_\mathrm{ST}\mathbf{w}_\mathrm{ST}^\mathrm{H}\mathbf{h}_\mathrm{TM}^\mathrm{H}$, 
and $c_\mathrm{M} \! \triangleq \! \mathbf{h}_\mathrm{TM}\mathbf{w}_\mathrm{ST}\mathbf{w}_\mathrm{ST}^\mathrm{H}\mathbf{h}_\mathrm{TM}^\mathrm{H}$.

\begin{algorithm}[t]
        \small
	\caption{BCD-PSO based scheme}
        \renewcommand{\algorithmicensure}{}
	\textbf{Input:} Initialize RIS's phase shift vector $\mathbf{v}^0=\mathbf{1}_N$; \\
	\textbf{Output:} $\mathbf{\Phi}^*, P_\mathrm{J}^*$;
	\label{alg1}
	\begin{algorithmic}[1]
		\STATE $r \leftarrow 0$;
            \WHILE{convergence condition is not satisfied \emph{and} $r < R$}
                \FOR{each RIS's reflective element $\ell \in \{1,2,\ldots,N\}$}
                    \STATE Update $\hat{\theta}_\ell^{r+1}$ according to  (\ref{theta_r+1});
                    \FOR{each particle $k \in \{1,2,\ldots,K\}$}
                        \STATE Initialize $\nu_k(0)\!=\!0, x_k(0)\!=\!o_{\mathrm{indiv},k}\!=\!o_{\mathrm{group}}\!=\!\hat{\theta}_\ell^{r\!+\!1}, F_{\mathrm{indiv},k}\!=\!F_{\mathrm{group}}=\mathrm{Inf}$;
                    \ENDFOR
                    \STATE $t \leftarrow 0$;
                    \WHILE{$t < T$}
                        \FOR{each particle $k \in \{1,2,\ldots,K\}$}
                            \STATE Update $\nu_k(t+1)$, $x_k(t+1)$ and $f_k(t+1)$ according to  (\ref{velocity}),  (\ref{position}) and  (\ref{fitness}), respectively;
                            \IF{$f_k(t+1) < F_{\mathrm{indiv},k}$}
                                \STATE $o_{\mathrm{indiv},k}=x_k(t+1), F_{\mathrm{indiv},k}=f_k(t+1)$;
                            \ENDIF
                            \IF{$f_k(t+1) < F_{\mathrm{group}}$}
                                \STATE $o_{\mathrm{group}}=x_k(t+1), F_{\mathrm{group}}=f_k(t+1)$;
                            \ENDIF
                        \ENDFOR
                        \STATE $t \leftarrow t+1$;
                    \ENDWHILE
                \ENDFOR
                \STATE $r \leftarrow r+1$;
            \ENDWHILE
            \STATE Calculate $\mathbf{\Phi}^*$ and $P_\mathrm{J}^*$ according to $\mathbf{\Phi}^*=\mathrm{diag}(\mathbf{v}^{r})$ and  (\ref{eq9}).
	\end{algorithmic}  
\end{algorithm}

\subsection{Procedures of the Proposed BCD-PSO}
Next, we apply spirits of BCD and PSO algorithms \cite{articleBCD,eberhart1995new} and tailor them to solve the problem. The proposed BCD-PSO is summarized in Algorithm \ref{alg1}, which is conducted in an iterative fashion. Within each round of iteration, we update each element's phase shift successively. When updating the $\ell$th element's phase shift, other elements' phase shifts remain unchanged. Specifically, the update of each phase shift consists of two steps: a \emph{Jamming Optimization} procedure and a \emph{Monitoring Optimization} procedure.

In the $r$th iteration, the RIS's phase shift vector is denoted by $\mathbf{v}^r \! \triangleq \! \mathrm{diag}([v_1^r ~ v_2^r ~ \ldots ~ v_N^r])$, $r=1,2,\ldots,R$, where $R$ is the maximum number of iteration within the BCD-PSO. 
The optimization problem of updating the $\ell$th phase shift is
\begin{align}
    \boldsymbol{P3} \ \min_{|v^r_\ell|^2=1} & \ \frac{\mathrm{Re}\{v^r_\ell\alpha^r_{\mathrm{ST},\ell}\} + \beta^r_{\mathrm{ST},\ell} - \sigma_\mathrm{SR}^2 \gamma_\mathrm{SR,th}}{\gamma_\mathrm{SR,th}[\mathrm{Re}\{v^r_\ell\alpha^r_{\mathrm{J},\ell}\} + \beta^r_{\mathrm{J},\ell}]}  \\
    \text{s.t.:} & \ \sigma_\mathrm{M}^2 \gamma_\mathrm{M,th} - \mathrm{Re}\{v^r_\ell\alpha^r_{\mathrm{M},\ell}\} - \beta^r_{\mathrm{M},\ell} \leq 0, 
\end{align}
where $\alpha^r_{\mathrm{X},\ell} \! \triangleq \! 2(b_{\mathrm{X},\ell} + a_{\mathrm{X},\ell} \sum_{m,m\neq \ell}{a_{\mathrm{X},m}^\mathrm{H}v_m^{r\mathrm{H}}})$, $\beta^r_{\mathrm{X},\ell} \! \triangleq \! |a_{\mathrm{X},\ell}|^2 + |\sum_{m,m\neq \ell}{v^r_ma_{\mathrm{X},m}}|^2 + 2\mathrm{Re}\{\sum_{m,m\neq \ell}{v^r_mb_{\mathrm{X},m}}\} + c_\mathrm{X}$, $a_{\mathrm{X},m}$ and $b_{\mathrm{X},m}$ are the $m$th elements in $\mathbf{a}_\mathrm{X}$ and $\mathbf{b}_\mathrm{X}$, respectively, and the subscript $\mathrm{X} \in \{\mathrm{ST,J,M}\}$. We further represent these complex-valued parameters with the amplitudes and phases, including $v^r_\ell \!= \!e^{j\theta^r_\ell}$, $\alpha^r_{\mathrm{ST},\ell} \!= \!\rho^r_{\mathrm{ST},\ell}e^{j\theta^r_{\mathrm{ST},\ell}}$, $\alpha^r_{\mathrm{J},\ell} \!= \!\rho^r_{\mathrm{J},\ell}e^{j\theta^r_{\mathrm{J},\ell}}$, and $\alpha^r_{\mathrm{M},\ell} \!= \!\rho^r_{\mathrm{M},\ell}e^{j\theta^r_{\mathrm{M},\ell}}$. Then, we get $\boldsymbol{P4}$:
\begin{align}
    \boldsymbol{P4} & \min_{0 \leq \theta^r_\ell \leq 2\pi} \frac{\rho^r_{\mathrm{ST},\ell}\cos(\theta^r_\ell \! + \! \theta^r_{\mathrm{ST},\ell}) \! + \! \beta^r_{\mathrm{ST},\ell} \! - \! \sigma_\mathrm{SR}^2 \gamma_\mathrm{SR,th}}{\gamma_\mathrm{SR,th}[\rho^r_{\mathrm{J},\ell}\cos(\theta^r_\ell+\theta^r_{\mathrm{J},\ell}) + \beta^r_{\mathrm{J},\ell}]} \label{eq5}\\
    & \text{s.t.:} \ \sigma_\mathrm{M}^2 \gamma_\mathrm{M,th} - \rho^r_{\mathrm{M},\ell}\cos(\theta^r_\ell+\theta^r_{\mathrm{M},\ell}) - \beta^r_{\mathrm{M},\ell}\leq 0. \label{eq7}
\end{align}
The BCD-PSO scheme are then detailed as follows.

\subsubsection{Jamming Optimization} \label{JO}

Based on  (\ref{eq5}), there is a finite number of extremums within $[0,2\pi]$. Taking the derivative of  (\ref{eq5}) with respect to $\theta^r_\ell$ and letting it be 0, we get $B^r + C^r\sin(\theta^r_\ell + \psi^r_\ell) = 0$, 
where $B^r \! \triangleq \! \gamma_\mathrm{SR,th}\rho^r_{\mathrm{ST},\ell}\rho^r_{\mathrm{J},\ell}\sin(\theta^r_{\mathrm{J},\ell} \! - \! \theta^r_{\mathrm{ST},\ell})$, $C^r \! \triangleq \! \sqrt{p^{r2} \! + \! q^{r2} \! - \! 2p^{r}q^{r}\cos(\theta^r_{\mathrm{ST},\ell} \! - \! \theta^r_{\mathrm{J},\ell})}$, $\psi^r_\ell \triangleq \mathrm{arctan}(\frac{p^{r}\sin(\theta^r_{\mathrm{J},\ell}) \! - \! q^{r}\sin(\theta^r_{\mathrm{ST},\ell})}{p^{r}\cos(\theta^r_{\mathrm{J},\ell}) \! - \! q^{r}\cos(\theta^r_{\mathrm{ST},\ell})})$ with $p^{r2} \triangleq \gamma_\mathrm{SR,th} \rho^r_{\mathrm{J},\ell}(\beta^r_{\mathrm{ST},\ell} \! - \! \sigma_\mathrm{SR}^2 \gamma_\mathrm{SR,th})$ and $q^{r2} \triangleq \rho^r_{\mathrm{ST},\ell} \beta^r_{\mathrm{J},\ell}$. Then, we have
\begin{eqnarray} \label{theta_r+1}
    \left\{
        \begin{array}{lcl}
            \theta^{r+1}_{\ell,1} = \mathrm{arcsin}\left(\frac{B^r}{C^r}\right) - \psi^r_\ell, \\
            \theta^{r+1}_{\ell,2} = \pi - \mathrm{arcsin}\left(\frac{B^r}{C^r}\right) - \psi^r_\ell.
        \end{array}
    \right.
\end{eqnarray}
Then, we substitute $\theta^{r+1}_{\ell,1}$ and $\theta^{r+1}_{\ell,2}$ into  (\ref{eq5}), respectively, and set a solution $\hat{\theta}^{r+1}_\ell$ equal to the one with smaller objective.

\subsubsection{Monitoring Optimization} \label{MO}

Next, we use PSO to search for a better solution. We first initializes $K$ particles. The $k$th particle, $k=1,2,\ldots,K$, is characterized by the iteratively updated position, velocity and fitness, denoted by $x_k(t)$, $\nu_k(t)$ and $f_k(t)$, respectively, where $t$ denotes the $t$-th iteration, $t=1,2,\ldots,T$, and $T$ is the maximum number of iteration. We define the position as element's phase shift. All particles will evolve differently under random rules, such that better solution can be discovered.

Specifically, the position's update is given by
\begin{equation} \label{position}
    x_k(t+1) = x_k(t) + \nu_k(t+1), 
\end{equation}
where $\nu_k(t)$ is the $k$th particle's updated velocity at the $t$-th iteration. Correspondingly, the velocity is obtained by
\begin{align} \label{velocity}
    \nu_k(t+1) &= w(t) \nu_k(t) + c_1 r_1(t) [o_{\mathrm{indiv},k} - x_k(t)] \nonumber \\
    &+ c_2 r_2(t) [o_{\mathrm{group}} - x_k(t)],
\end{align}
where $w(t)$ is the inertia weight at the $t$-th iteration, to dampen the velocity. Lower inertia weight confines the search within a relatively local range, and vice verse. To effectively attain the convergence, we let $w(t)$ decrease as $t$ increases, where $w(t)=w_\mathrm{max} - (\frac{w_\mathrm{max}-w_\mathrm{min}}{T})t$, with the pre-set upper limit $w_\mathrm{max}$ and lower limit $w_\mathrm{min}$. Here, $o_{\mathrm{indiv},k}$ and $o_{\mathrm{group}}$ represent current optimized individual and group positions, $c_1$ and $c_2$ are individual and group learning coefficients that control the influence of $o_{\mathrm{indiv},k}$ and $o_{\mathrm{group}}$, respectively, and $r_1(t)$ and $r_2(t)$ both are uniformly-distributed random numbers within $[0,1]$ to generate difference across evolvement of each particle. 

In PSO algorithm, fitness is calculated after updating position. We define fitness according to Eqs.~(\ref{eq5}) and (\ref{eq7}) as
\begin{align} \label{fitness}
    &f_k(t+1) = \frac{\rho^r_{\mathrm{ST},\ell}\cos(x_k(t \! + \! 1) \! + \! \theta^r_{\mathrm{ST},\ell}) \! + \! \beta^r_{\mathrm{ST},\ell} \! - \! \sigma_\mathrm{SR}^2 \gamma_\mathrm{SR,th}}{\gamma_\mathrm{SR,th}[\rho^r_{\mathrm{J},\ell}\cos(x_k(t+1)+\theta^r_{\mathrm{J},\ell}) + \beta^r_{\mathrm{J},\ell}]} \nonumber \\
    & \! + \! \Xi\{\sigma_\mathrm{M}^2 \gamma_\mathrm{M,th} \! - \! \rho^r_{\mathrm{M},\ell}\cos(x_k(t \! + \! 1) \! + \! \theta^r_{\mathrm{M},\ell}) \! - \! \beta^r_{\mathrm{M},\ell} \leq 0\},
\end{align}
where $\Xi\{L\} \triangleq (1-L)c_\mathrm{p}$, $L \in \{0,1\}$ is a bool variable, and $c_\mathrm{p}$ is a large positive value for violating  (\ref{eq7}). Moreover, the optimized fitnesses corresponding to $o_{\mathrm{indiv},k}$ and $o_{\mathrm{group}}$ are denoted by $F_{\mathrm{indiv},k}$ and $F_{\mathrm{group}}$, respectively. If $f_k(t\!+\!1) \! < \! F_{\mathrm{indiv},k}$, the $x_k(t\!+\!1)$ will be set to $o_{\mathrm{indiv},k}$. Furthermore, if $f_k(t\!+\!1) \! < \! F_{\mathrm{group}}$, update the $o_{\mathrm{group}}$ by $x_k(t\!+\!1)$.

Then, we iteratively update phase shift of each element in sequence until the maximum number is reached or the convergence condition ($\epsilon = \frac{\sum_{n=1}^N{\|\mathbf{v}_n^{r+1}-\mathbf{v}_n^{r}\|}}{\sum_{n=1}^N{\|\mathbf{v}_n^{r}\|}} \leq \epsilon_\mathrm{th}$) is triggered, 
where $\epsilon_\mathrm{th}$ is the specified threshold. According to the above steps, we can get the suboptimal solution of RIS's phase shift matrix $\mathbf{\Phi}^*$ and LJs's transmission power $P_\mathrm{J}^*$, respectively.

\section{Simulation Evaluations} \label{S5}
In this section, we present simulation results to evaluate performances of our proposed BCD-PSO based scheme. All results are averaged over 200 independent simulations, and the simulation parameters are given below unless otherwise stated.


We configure a three-dimensional coordinate as shown in Fig.~\ref{figure1}, where RIS is placed at $(0,20,3)$ (unit: m). ST and SR are located at $(20,0,3)$ and $(20,100,0)$, respectively. LM and LJs are randomly distributed in the area centered at $(20,150,0)$ with a radius of $20~\mathrm{m}$. We assume that $\mathbf{h}_\mathrm{TR}$, $\mathbf{h}_\mathrm{TM}$, and $\mathbf{h}_\mathrm{KR}$ follow the Rayleigh-fading, and RIS-related links follow the Rician fading. Rician factors of $\mathbf{H}_\mathrm{TI}$ and $\mathbf{h}_\mathrm{IR}$ both are $2$ due to obstacles, and Rician factors of $\mathbf{h}_\mathrm{IM}$ and $\mathbf{H}_\mathrm{KI}$ both equal to $10$. The large-scale path loss is modeled by $\mathrm{PL} = \mathrm{PL}_0 - 10 \mu \mathrm{log}(\frac{d}{d_0})$, where $\mathrm{PL}_0$ is the path loss at reference distance $d_0=1~\mathrm{m}$, and $\mu$ is the path loss factor. Due to scatters, path loss factors of $\mathbf{h}_\mathrm{TR}$, $\mathbf{h}_\mathrm{IR}$, $\mathbf{h}_\mathrm{KR}$, and $\mathbf{h}_\mathrm{TM}$ are $4$, and path loss factors of $\mathbf{H}_\mathrm{TI}$, $\mathbf{H}_\mathrm{KI}$, and $\mathbf{h}_\mathrm{IM}$ are set to $2$ with the selection of RIS's location. The noise power at SR and LM both are $-90~\mathrm{dBm}$. Moreover, $\gamma_\mathrm{M,th}$ equal to $12~\mathrm{dB}$, $\epsilon_\mathrm{th}$ is $0.01$, $w_\mathrm{min}$ and $w_\mathrm{max}$ are set to $0$ and $2\pi$, respectively, and iteration numbers $R$ and $T$ are $50$ and $80$, respectively.

To demonstrate performance improvement achieved by the proposed BCD-PSO, we define successfully monitoring probability (SMP) and successfully jamming probability (SJP) as the proportion of successfully monitoring and jamming on suspicious communications in simulations, respectively. Moreover, we consider the following comparative schemes:
\begin{itemize} 


\item{\emph{PSO Scheme}: Only PSO-based monitoring optimization procedure is adopted, and initial positions of particles are randomly selected within $[0,2\pi]$.}

\item{\emph{Simulated Annealing (SA) Scheme}: Compared to the \emph{PSO Scheme}, SA algorithm \cite{4358775} is used to optimize $\mathbf{\Phi}$ and $P_\mathrm{J}$.}

\item{\emph{PSO-Domain Scheme}: Compared to \emph{PSO Scheme}, initial positions of particles are randomly selected within the feasible domains given by  (\ref{eq7}).}

\item{\emph{BCD-Domain Scheme}: Only the BCD-based jamming optimization procedure is employed under the feasible domains given by  (\ref{eq7}).}

\item{\emph{BCD-SA Scheme}: It uses BCD-based jamming optimization procedure and adopts SA algorithm in monitoring optimization procedure.}

\item{\emph{Random Phase}: The phase shift of each RIS's reflective element is independently generated from 0 to $2\pi$.}

\item{\emph{Without RIS}: There is no RIS in the system, which means that all CSI for RIS-related links are set to zero matrices.}
\end{itemize}

\begin{figure*}[!t]
\vspace*{-0.3\baselineskip}
    \centering
    \subfloat[]{\includegraphics[width=1.8in]{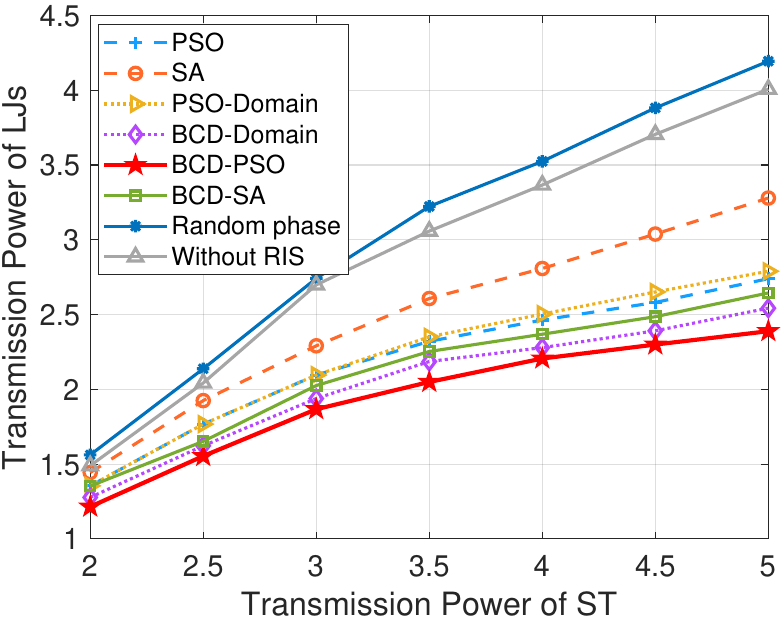}%
    \label{fig1.1}}
    \hfil
    \subfloat[]{\includegraphics[width=1.8in]{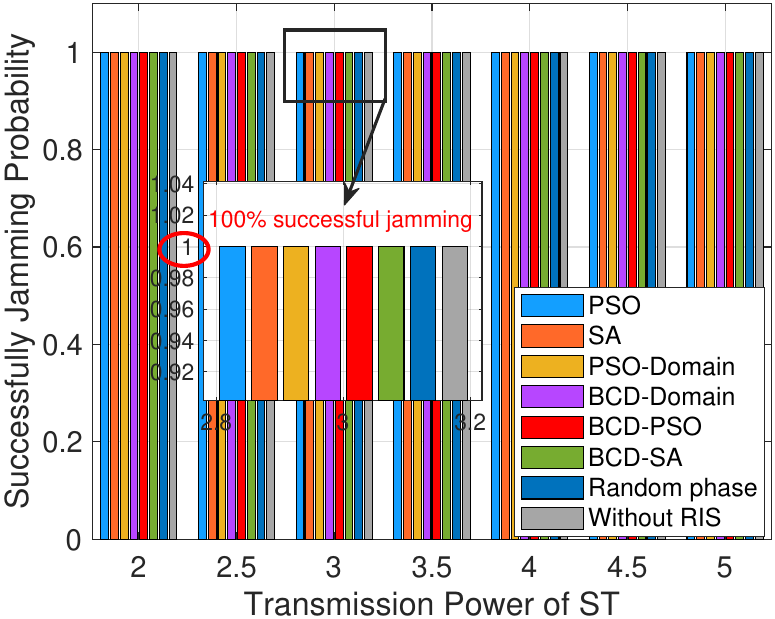}%
    \label{fig1.2}}
    \hfil
    \subfloat[]{\includegraphics[width=1.8in]{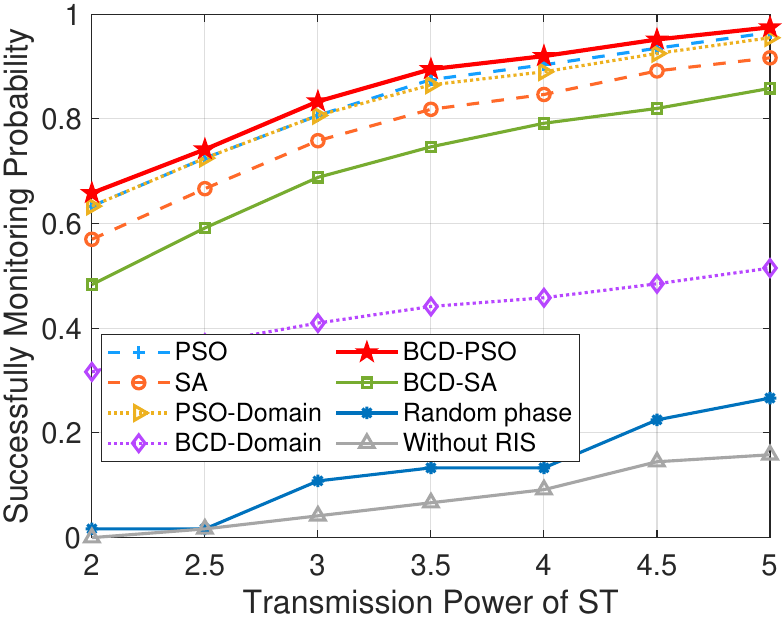}%
    \label{fig1.3}}
    \caption{Performances versus ST's transmission power with $K=6$, $N=10$, and $\gamma_\mathrm{SR,th}=-10\mathrm{dB}$. (a) LJs' transmission power. (b) SJP. (c) SMP.}
    \label{fig1}
\vspace*{-0.5\baselineskip}
\end{figure*}

In Fig.~\ref{fig1}, we study the effect of ST's transmission power on LJs' transmission power, SJP and SMP. It can be observed from Figs.~\ref{fig1}(a) and (b) that the proposed BCD-PSO achieve the lowest transmission power while attaining $100\%$ SJP, indicating completely block suspicious communications. The BCD-PSO also obtains the highest SMP according to Fig.~\ref{fig1}(c). By comparison, other schemes cannot simultaneously minimize LJs' transmission power and monitor suspicious communications. For example, the BCD-Domain has similar performance in resource utilization to the BCD-PSO, but its monitoring performance is much worse; the PSO can achieve similar SMP to the proposed BCD-PSO, but LJs would consume more transmission power; and both Random Phase and Without RIS cannot realize effective monitoring and resource saving. Moreover, as ST's transmission power increases, LJs need more power and LM's monitoring performance gets better.

\begin{figure}[!t]
\vspace*{-0.3\baselineskip}
    \centering
    \subfloat[]{\includegraphics[width=1.72in]{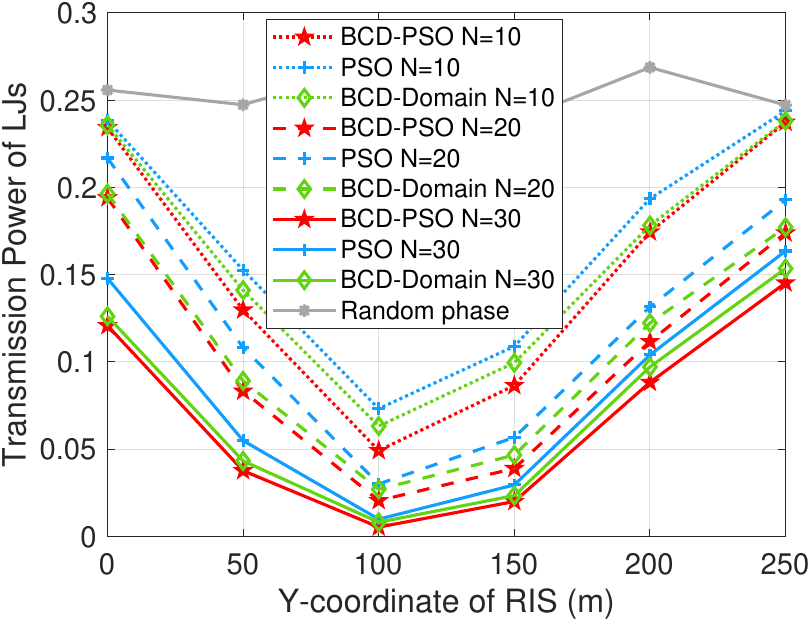}%
    \label{fig2.1}}
    \hfil
    \subfloat[]{\includegraphics[width=1.7in]{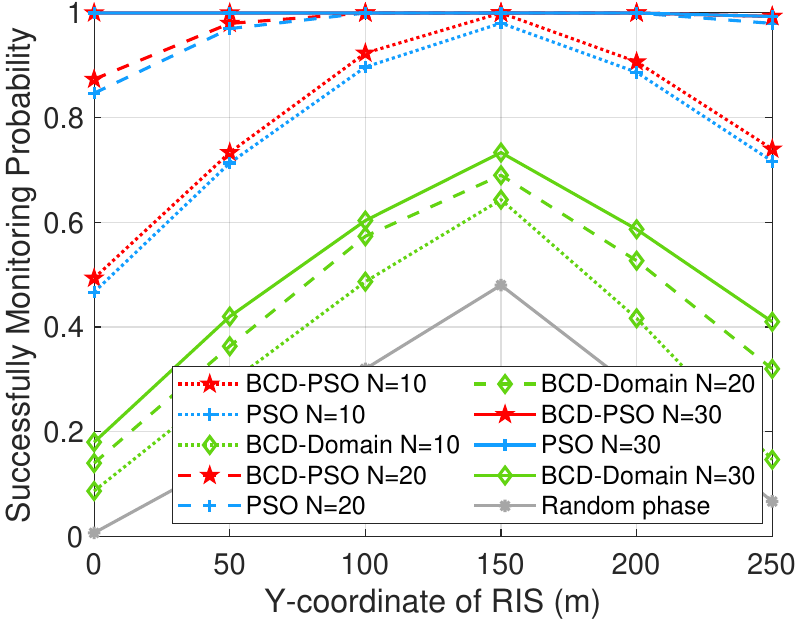}%
    \label{fig2.2}}
    \caption{Performances versus Y-coordinate of RIS with $K=8$, $\gamma_\mathrm{SR,th}=-10\mathrm{dB}$, and $P_\mathrm{ST}=1$. (a) LJs' transmission power. (b) SMP.}
    \label{fig2}
\vspace*{-0.5\baselineskip}
\end{figure}

Fig.~\ref{fig2} shows variations of LJs' transmission power and SMP with RIS's location and the number of elements in the case of 100\% SJP. From the results, transmission power of LJs decreases with the smaller distance between RIS and SR, while SMP increases with decrement of the distance between RIS and LM. Therefore, the preferred location of RIS is between SR and LM, that is, RIS's Y-coordinate should be confined within $[100,150]$ under the environment set for simulations. It can also be observed that LJs' transmission power decreases and SMP increases when the number of elements gets larger, which demonstrates the crucial role the RIS plays. Furthermore, compared among the BCD-PSO, PSO and BCD-Domain schemes, the PSO tends to optimize monitoring performance, the BCD-Domain tends to reduce LJs' transmission power, and the proposed BCD-PSO can achieve better performances in both terms of monitoring and resource utilization.

\begin{figure}[!t]
\vspace*{-0.3\baselineskip}
    \centering
    \subfloat[]{\includegraphics[width=1.7in]{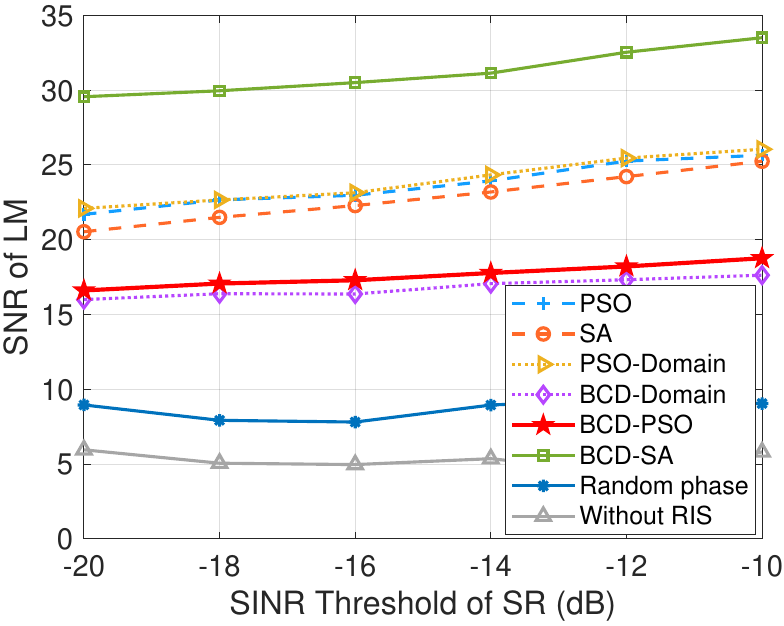}%
    \label{fig3.1}}
    \hfil
    \subfloat[]{\includegraphics[width=1.72in]{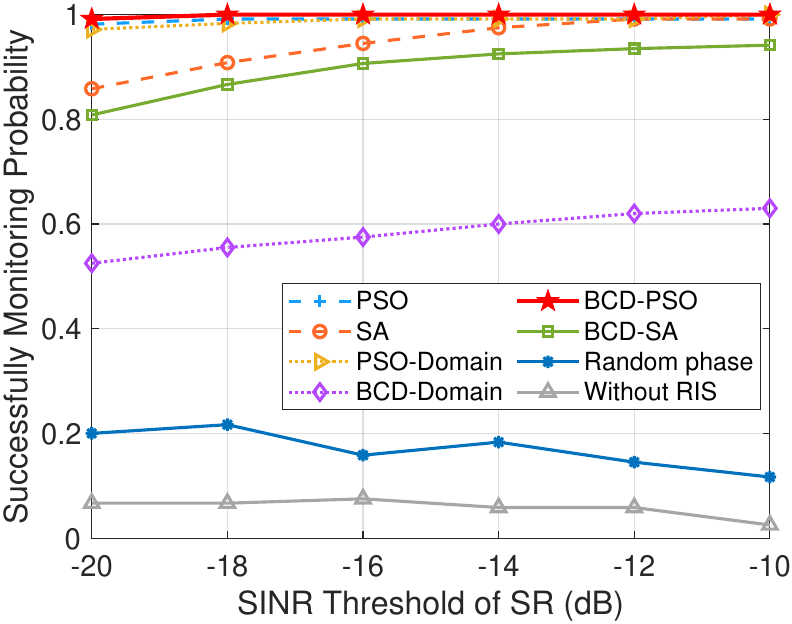}%
    \label{fig3.2}}
    \caption{Performances versus $\gamma_\mathrm{SR,th}$ with $K=6$, $N=20$, and $P_\mathrm{ST}=3$. (a) SNR of LM. (b) SMP.}
    \label{fig3}
\vspace*{-0.5\baselineskip}
\end{figure}

As shown in Fig.~\ref{fig3}, we compare the average SNR of LM with the SMP under different schemes. It is noted that LM's SNR obtained by the BCD-PSO is not the highest one. However, the BCD-PSO can attain the highest SMP among comparative schemes, with completely near-$100\%$ SMP. The results above indicate that the BCD-PSO is more robust to uncertainties in wireless communications environment.

\section{Conclusions} \label{S6}
In this paper, a RIS-assisted legitimate monitoring and jamming system towards industrial environment was studied, in which LM and LJs conduct legitimate monitoring and jamming on suspicious communications, respectively. To minimize the transmission power of LJs while preventing SR from successful decoding and ensuring continuous monitoring, we proposed a robust BCD-PSO based scheme. We sequentially updated phase shift of each reflective elements and took results of the \emph{Jamming Optimization} procedure as initial values of the \emph{Monitoring Optimization} procedure. Simulation results showed that our proposed BCD-PSO is able to achieve better performances in terms of LJs' transmission power, SMP and robustness. We can also effectively optimize the deployment of RIS towards different objectives. 




\bibliographystyle{IEEEtran}
\bibliography{reference}

\begin{thebibliography}{10}
\providecommand{\url}[1]{#1}
\csname url@samestyle\endcsname
\providecommand{\newblock}{\relax}
\providecommand{\bibinfo}[2]{#2}
\providecommand{\BIBentrySTDinterwordspacing}{\spaceskip=0pt\relax}
\providecommand{\BIBentryALTinterwordstretchfactor}{4}
\providecommand{\BIBentryALTinterwordspacing}{\spaceskip=\fontdimen2\font plus
\BIBentryALTinterwordstretchfactor\fontdimen3\font minus \fontdimen4\font\relax}
\providecommand{\BIBforeignlanguage}[2]{{%
\expandafter\ifx\csname l@#1\endcsname\relax
\typeout{** WARNING: IEEEtran.bst: No hyphenation pattern has been}%
\typeout{** loaded for the language `#1'. Using the pattern for}%
\typeout{** the default language instead.}%
\else
\language=\csname l@#1\endcsname
\fi
#2}}
\providecommand{\BIBdecl}{\relax}
\BIBdecl

\bibitem{10128963}
H.~Jing, W.~Cheng, and X.-G. Xia, ``Fast transceiver design for {RIS}-assisted {MIMO} mmwave wireless communications,'' \emph{IEEE Trans. Wireless Commun.}, vol.~22, no.~12, pp. 9939--9954, Dec. 2023.

\bibitem{10061580}
T.~Yin, L.~Li, W.~Lin, H.~Hu, D.~Ma, J.~Liang, T.~Bai, C.~Pan, and Z.~Han, ``Joint active and passive beamforming optimization for multi-{IRS}-assisted wireless communication systems: A covariance matrix adaptation evolution strategy,'' \emph{IEEE Trans. Veh. Technol.}, vol.~72, no.~7, pp. 9281--9292, Jul. 2023.

\bibitem{9446526}
H.~Niu, Z.~Chu, F.~Zhou, Z.~Zhu, M.~Zhang, and K.-K. Wong, ``Weighted sum secrecy rate maximization using intelligent reflecting surface,'' \emph{IEEE Trans. Commun.}, vol.~69, no.~9, pp. 6170--6184, Sept. 2021.

\bibitem{pan2022overview}
C.~Pan, G.~Zhou, K.~Zhi, S.~Hong, T.~Wu, Y.~Pan, H.~Ren, M.~Di~Renzo, A.~L. Swindlehurst, R.~Zhang \emph{et~al.}, ``An overview of signal processing techniques for {RIS/IRS}-aided wireless systems,'' \emph{IEEE J. Sel. Top. Signal Process}, vol.~16, no.~5, pp. 883--917, Aug. 2022.

\bibitem{9672153}
Y.~Sun, K.~An, J.~Luo, Y.~Zhu, G.~Zheng, and S.~Chatzinotas, ``Outage constrained robust beamforming optimization for multiuser {IRS}-assisted anti-jamming communications with incomplete information,'' \emph{IEEE Internet of Things J.}, vol.~9, no.~15, pp. 13\,298--13\,314, Aug. 2022.

\bibitem{10066528}
S.~Arzykulov, A.~Celik, G.~Nauryzbayev, and A.~M. Eltawil, ``Artificial noise and {RIS}-aided physical layer security: Optimal {RIS} partitioning and power control,'' \emph{IEEE Wireless Commun. Lett.}, vol.~12, no.~6, pp. 992--996, Jun. 2023.

\bibitem{9685957}
V.~Kumar, M.~F. Flanagan, D.~W. Kwan~Ng, and L.-N. Tran, ``On the secrecy rate under statistical {QoS} provisioning for {RIS}-assisted {MISO} wiretap channel,'' in \emph{IEEE Global Communications Conference}, Madrid, Spain, 2021.

\bibitem{9262884}
Z.~Chu, W.~Hao, P.~Xiao, D.~Mi, Z.~Liu, M.~Khalily, J.~R. Kelly, and A.~P. Feresidis, ``Secrecy rate optimization for intelligent reflecting surface assisted {MIMO} system,'' \emph{IEEE Trans. Inf. Forensics Secur.}, vol.~16, pp. 1655--1669, 2021.

\bibitem{9417411}
L.~Sun, Y.~Zhang, and A.~L. Swindlehurst, ``Wireless physical-layer surveillance via proactive eavesdropping and alternate jamming,'' in \emph{IEEE Wireless Communications and Networking Conference (WCNC)}, Nanjing, China, 2021.

\bibitem{9797949}
M.~Wang, Q.~Du, and L.~Zhang, ``Minimum-{SNR} maximization for robust {IRS}-assisted legitimate monitoring system,'' in \emph{IEEE Conference on Computer Communications Workshops}, New York, NY, USA, 2022.

\bibitem{hu2022analysis}
G.~Hu, Z.~Li, J.~Si, K.~Xu, Y.~Cai, D.~Xu, and N.~Al-Dhahir, ``Analysis and optimization of {STAR-RIS}-assisted proactive eavesdropping with statistical {CSI},'' \emph{IEEE Trans. Veh. Technol.}, vol.~72, no.~5, pp. 6850--6855, May 2022.

\bibitem{9779086}
Y.~Sun, K.~An, Y.~Zhu, G.~Zheng, K.-K. Wong, S.~Chatzinotas, H.~Yin, and P.~Liu, ``{RIS}-assisted robust hybrid beamforming against simultaneous jamming and eavesdropping attacks,'' \emph{IEEE Trans. Wireless Commun.}, vol.~21, no.~11, pp. 9212--9231, Nov. 2022.

\bibitem{10333823}
L.~Zhang, Q.~Du, L.~Lu, and S.~Zhang, ``A low-complexity estimation scheme for separated reflecting channels of {RIS}-assisted {MIMO} systems towards extended coverage,'' in \emph{IEEE 98th Vehicular Technology Conference}, Hong Kong, 2023.

\bibitem{articleBCD}
Y.~Xu and W.~Yin, ``A block coordinate descent method for regularized multiconvex optimization with applications to nonnegative tensor factorization and completion,'' \emph{SIAM J. Imaging Sci.}, vol.~6, no.~3, pp. 1758--1789, 2013.

\bibitem{eberhart1995new}
R.~Eberhart and J.~Kennedy, ``A new optimizer using particle swarm theory,'' in \emph{Proceedings of the Sixth International Symposium on Micro Machine and Human Science}, 1995, pp. 39--43.

\bibitem{4358775}
S.~Bandyopadhyay, S.~Saha, U.~Maulik, and K.~Deb, ``A simulated annealing-based multiobjective optimization algorithm: {AMOSA},'' \emph{IEEE Trans. Evol. Comput.}, vol.~12, no.~3, pp. 269--283, Jun. 2008.

\end{thebibliography}

\end{document}